\documentclass[letterpaper]{article} 
\usepackage{aaai24}  
\usepackage{times}  
\usepackage{helvet}  
\usepackage{courier}  
\usepackage[hyphens]{url}  
\usepackage{graphicx} 
\usepackage{natbib}  
\usepackage{caption} 
\usepackage{algorithm}
\usepackage{algorithmic}
\usepackage{tabularray}
\usepackage{pdflscape}

\usepackage{newfloat}
\usepackage{listings}
\DeclareCaptionStyle{ruled}{labelfont=normalfont,labelsep=colon,strut=off} 
\floatstyle{ruled}
\newfloat{listing}{tb}{lst}{}
\floatname{listing}{Listing}
\title{Unsocial Intelligence: an Investigation of the Assumptions of AGI Discourse}
\author{
    Borhane Blili-Hamelin\equalcontrib \textsuperscript{\rm 1},
    Leif Hancox-Li\equalcontrib \textsuperscript{\rm 2},
    Andrew Smart\textsuperscript{\rm 3}
}
\affiliations{
    \textsuperscript{\rm 1}AI Risk and Vulnerability Alliance\\

     \textsuperscript{\rm 2}vijil\\

     \textsuperscript{\rm 3}Google Research\\

}

\begin{document}

\maketitle

\begin{abstract}
Dreams of machines rivaling human intelligence have shaped the field of AI since its inception. Yet, the very meaning of human-level AI or artificial general intelligence (AGI) remains elusive and contested. Definitions of AGI embrace a diverse range of incompatible values and assumptions. Contending with the fractured worldviews of AGI discourse is vital for critiques that pursue different values and futures. To that end, we provide a taxonomy of AGI definitions, laying the ground for examining the key social, political, and ethical assumptions they make. We highlight instances in which these definitions frame AGI or human-level AI as a technical topic and expose the value-laden choices being implicitly made. Drawing on feminist, STS, and social science scholarship on the political and social character of intelligence in both humans and machines, we propose contextual, democratic, and participatory paths to imagining future forms of machine intelligence. The development of future forms of AI must involve explicit attention to the values it encodes, the people it includes or excludes, and a commitment to epistemic justice.
\end{abstract}

\section{Introduction}

There is no agreed-upon definition of artificial general intelligence (AGI) \cite{morris_levels_2023, mitchell_debates_2024}. Yet from influential AI companies \cite{deepmind_about_2022, openai_2018, meaker_aleph_meet_nodate, meta_agi} and AI researchers \cite{aguera_y_arcas_artificial_2023, openai_2018, morris_levels_2023, chollet_measure_2019, bubeck_sparks_2023} to the increasingly public worry \cite{future_of_life_institute_pause_nodate, center_for_ai_safety_statement_nodate} about existential risks \cite{bostrom_superintelligence_2014, mclean_risks_2023, naude_artificial_nodate, noy_four_2022}, AGI and human-level AI have become one of the dominant ways to imagine the future potential of AI. At the same time, AGI has had its own skeptics \cite{lecun, forbes, agi_techreview, Gebru_Torres_2024} who risk taking the target of their critiques to be more homogenous and congruent than it actually is. 

The lack of homogeneity in current conceptions of AGI is not a bug. It is a feature of the underlying topic: what might it mean for machines to have intelligence that rivals human intelligence? Substantive disagreements come with the value-laden character of both intelligence and of AI as a technology. By value-laden, we mean that political, social, and ethical values can and should shape conceptions of intelligence, technology, and their intersection.\cite{blili-hamelin_borhane_making_2023} We lack consensus on the definition of intelligence because of conflicting values. The same applies to the current lack of consensus on AGI.

Ruha Benjamin warns that ``a narrow definition of what even counts as technology or intelligence'' threatens the ability of communities to imagine worlds worth building.\cite{benjamin_imagination_2024} What forms of intelligence and technology are worth imagining are political and social questions. However, current approaches to AGI risk mistaking these questions for technical questions. We lay the groundwork for examining these potential mistakes by analyzing and classifying contingent assumptions underlying different definitions of AGI. Contending with the fractured worldviews among conceptions of AGI plants the seeds for resisting harmfully narrow conceptions of machine intelligence. 

Our paper proceeds as follows. We begin by providing a framework for thinking of AGI as inheriting the value-laden features of both intelligence and technology (Section ~\ref{human}). We then investigate the value-laden, contingent choices made by influential accounts of AGI and human-level AI (Section ~\ref{choices}). Rather than framing these choices as inherently misguided, we see them as opportunities to investigate the limited but heterogeneous range of questions currently being asked about would-be human-level AI. Finally, we sketch pathways for more contextual, democratically legitimate, and participatory perspectives on what forms of machine intelligence are worth imagining (Sections \ref{towards} and \ref{democracy}).

Our investigation is not limited to accounts using the exact phrase “artiﬁcial general intelligence”. Like Morris et al. 2023, our topic is the long history of treating ``[a]chieving human-level ``intelligence'' as the ``north-star goal'' of the AI field \cite{morris_levels_2023}, dating at least as far back as the 1955 Dartmouth AI Conference \cite{mccarthy_proposal_1955}. Discussions of ``human-level AI'', ``general AI'', or ``AI'' (such as in the phrase ``strong AI'' \cite{searle_minds_1980}) fall within the scope of our account.

\section{Between human intelligence and technology: AGI's dual value-laden pedigrees}
\label{human}

Building on STS scholarship on the values embedded in technology \cite{winner_artifacts_1980}, research communities like FAccT, AIES, and CHI have taken a deep interest in the political, social, and ethical values embedded both in AI tools and in the practices that surround AI \cite{fishman_should_2022, dotan_value-laden_2020, birhane_values_2022, scheuerman2021datasets, hutchinson_evaluation_2022, bommasani_evaluation_2022, denton_bringing_2020, denton2021genealogy, mathur_disordering_2022, shilton_values_2018, broussard_artificial_2019, green_data_2021, blodgett_language_2020, viljoen_relational_2021, abebe_roles_2020, birhane_towards_2021, blili-hamelin_borhane_making_2023, costanza-chock_design_2020}. AGI and human-level AI concern not only existing technologies, but also the technologies that many AI builders, researchers, and organizations dream of building in the future. When companies describe their official ``long term aim [as] to solve intelligence, developing more general and capable problem-solving systems, known as artificial general intelligence (AGI)'' \cite{deepmind_about_2022}, or when thought leaders claim that ``[m]itigating the risk of extinction from AI should be a global priority alongside other societal-scale risks such as pandemics and nuclear war'' \cite{center_for_ai_safety_statement_nodate}, they engage in ``the process of negotiating between competing perspectives, values, and goals'' \cite{green_data_2021}. 

We interrogate the political, social, and ethical questions undergirding AGI discourse by paying attention to how different accounts embody competing visions for the future of technology. We don't see value-laden assumptions as a flaw. We see them as choices that admit legitimate disagreement through competing values. Our account of alternative paths forward (Section \ref{towards}) strives to be reflective and explicit about the political and social assumptions of the visions we call for---such as democracy, epistemic justice, contextualism, and participation. Throughout, we also embrace the value of reflectiveness \cite{boyarskaya_overcoming_2020, prunkl_institutionalizing_2021} about political, social, and ethical assumptions. We believe that making value assumptions explicit to ourselves and to others often makes for better individual and collective decisions---we might say, for more intelligent ``experiments in living'' \cite{mill_liberty_2003, anderson_john_1991}. Our aim in emphasizing reflectiveness is to point to more legitimate methods of determining the values that we want in AI. When large corporations are the only entities championing the narrative that their visions of AGI will transform everyone's lives, what and whose values ultimately win out becomes a question of power divorced from legitimacy \cite{burrell_introduction_2024}.

In the rest of this section, we examine a second and more often overlooked root of AGI's value-laden character: \emph{intelligence}. Debates about defining and measuring human intelligence are crucial to anticipating the value-laden aspects of AGI for at least three reasons. Firstly, many of the reasons for why definitions of human intelligence are value-laden carry over to the case of AI. Secondly, some attempts at defining AGI use concepts or methods from human intelligence research. Thirdly, the question of whether machines match or surpass human intelligence faces the challenge of specifying what counts as human intelligence and of evaluating human intelligence.

\subsection{Intelligence is value-laden because it is a thick concept}
Intelligence is what philosophers would call a \emph{thick evaluative concept}: it includes both descriptive and normative elements \cite{anderson_situated_2002, kirchin_thick_2013, kirchin_thick_2017}. It contains descriptive elements about what empirical phenomena fall under the concept of intelligent behavior. It also contains a normative element. Evaluating intelligence inevitably involves assessing the desirability of specific behaviors. Previous research \cite{anderson_situated_2002, alexandrova_democratising_2022, blili-hamelin_borhane_making_2023, cave_problem_2020} has argued that when we try to define or measure thick concepts, we inevitably embed ethical values in the design decisions we make about the boundaries of the concepts and the measurement methodology. Other examples of thick concepts that have value-laden definitions are health and well-being \cite{alexandrova_democratising_2022}. The fact that intelligence is value-laden is at the root of debates over its definition and validity.

The validity of the concept of general intelligence in humans has been extensively criticized. One of the principle targets of these critiques is Spearman's \emph{g}: a hidden, not directly observable factor expressing ``shared variance across a set of intercorrelating cognitive tasks'' \cite{warne_spearmans_2019}. Here we recap a few of the main critiques to provide some background for our later points about how some definitions of AGI make moves that are subject to similar critiques.

\subsubsection{History of ableism and racism}
\label{ableism}

General intelligence metrics for humans have been criticized as ableist and racist \cite{anderson_situated_2002, summerfield_natural_2023}. Anti-racist critiques of human intelligence tests cite the influence of teacher expectations and stereotype threat, which affect test-takers of different demographics differently \cite{osborne2001testing}. Disability advocates have argued that conventional standardized tests for intelligence are ableist because they do not consider how disabled individuals can perform within a broader social context \cite{pellicano_annual_2022}. In general, these critiques of \emph{g} draw on the fact that human experiences and behaviors are diverse and contextual in ways that are hard to capture in standardized tests \cite{legault2021neurodiversity}.

\subsubsection{Weaknesses of factor analysis}
\label{factor-analysis}

The use of factor analysis to discover \emph{g} as a causal factor has been extensively critiqued \cite{glymour_what_1998, johnson_realism_2016}. These critics argue that factor analysis, on its own, does not rule out alternative causal structures that could explain the same empirical results. Factor analysis can show that a common cause is a possible explanation of observed statistical patterns. But those same patterns could be explained by many other causal hypotheses.

We mention this critique of factor analysis not just to situate the epistemic status of human intelligence metrics, but also as background for our later discussion (Section \ref{g-ai}) of attempts in AI to find the equivalent of \emph{g} in AI systems. These attempts are subject to the same critique that factor analysis does not rule out alternative causal hypotheses.

One possible response to this critique of factor analysis is to treat \emph{g} in a deflationary way: to say that it is not explanatory but is merely a shorthand for the correlation between performances on multiple cognitive tests. In this way, \emph{g} could serve as a shorthand for \emph{communicating} performance on a group of tests, without committing to it being a real causal factor. This is analogous to formulations of AGI that strive for a deflationary approach, as we describe in Section \ref{deflate}.

\subsubsection{Circularity}
\label{circular}

Another critique of psychometrics work purporting to quantify human intelligence is that it cannot be validated without reference to intelligence tests, making such measurements circular \cite{summerfield_natural_2023, boring_intelligence_1923, richardson2017genes, popper2013knowledge}. As \citet{boring_intelligence_1923} put it over 100 years ago, ``[I]ntelligence is what the tests test''. \citet{boring_intelligence_1923}'s worry is not simply that intelligence tests need to be appealed to in validating intelligence tests. Rather, it is that the theoretical construct they purport to measure is itself motivated by setting aside the broad range of connotations of intelligence in favor of focusing on whatever can be measured through intelligence tests \cite{boring_intelligence_1923}. This critique points to how psychometrics research is insufficiently motivated and under-theorized. \citet{warne_spearmans_2019} is an example of recent work that doubles down on disconnecting research on \emph{g} from any attempt at anchoring psychometrics research in a substantive theoretical conception of intelligence. As with the critique of factor analysis (\ref{factor-analysis}),  deflationary approaches that avoid ascribing an explanatory role to \emph{g} may be a potential response to this worry. \footnote{For an overview of recent approaches to motivating psychomterics research, see \citet{deary_intelligence_2012}.}

\subsection{Relevance to AGI}

The critiques of \emph{g} and the recognition of intelligence as a thick concept provide the background for why the definition and measurement of human intelligence is a political and social question. For similar reasons, defining and measuring AGI is also a political and social question. AGI is also a thick concept involving both descriptive (what tasks or abilities fall under its definition) and normative (what counts as \emph{good} machine behavior) criteria. And, as we argue next, current definitions of AGI embed political and social values in ways that are often not explicitly acknowledged by their authors. In the next section, we identify the design choices that formulations of AGI make, emphasizing their contingency and the types of values that each choice embeds.

\section{The motley choices of AGI discourse}
\label{choices}

In this section, we look at some of the typical moves made when defining AGI. Our central contention is that these definitions always depend on assumptions about what is valuable to some group of people or what goals are worthwhile---whether implicitly or explicitly. This is expected once we understand that the topic of AGI and human-level AI sits at the intersection of two fundamentally value-laden questions. The topic of intelligence is fundamentally value-laden due to its nature as a thick concept (see Section \ref{human}). Moreover, the topic of what technologies and tools are \emph{worth building} is itself fundamentally value-laden \cite{costanza-chock_design_2020, birhane_values_2022, dotan_value-laden_2020}. No conception of AGI can escape the political, social, and ethical priorities that \emph{do} and \emph{should} shape any answer to the question of what is worth building, by whom, and for what purposes. Throughout this section, we highlight value-laden aspects of AGI discourse that pertain both to its character as a technology and to the nature of the task of defining intelligence. 

Some accounts of AGI attempt to adopt a kind of neutral stance. They may \emph{start} by declaring that they intend to have a deflationary, value-neutral take on AGI (similar to the deflationary view on \emph{g} we sketched in Section \ref{factor-analysis}), but even these ultimately end up going beyond a purely deflationary view (see Section \ref{fake-deflationary}). We start by sketching the deflationary view (Section \ref{deflate}), then proceed to look at some of the major AGI definitions out there and identify dimensions on which different definitions make different value-laden assumptions (Section \ref{dimensions-taxonomy}).

\subsection{Purportedly deflationary accounts of AGI}
\label{deflate}

Some definitions of AGI start out with seemingly neutral goals, such as creating a shared language or common standards for measuring AGI, without (at least at first) specifying any normative goals beyond that. This can be viewed as similar to the deflationary view of \emph{g} described in Section \ref{factor-analysis}: where \emph{g} is simply a correlational factor that you can measure from a battery of tests and does not necessarily represent any underlying causal or explanatory factor. One can adopt standards for the sake of being able to consistently compare systems against one another within those standards without assuming that those standards are the only ``real'' or ``objective'' standards.

An example of the deflationary view is represented (at least initially) in ``Levels of AGI'' \cite{morris_levels_2023}. At the start of the paper, the authors claim that they are seeking a ``common language to compare models, assess risks, and measure progress along the path to AGI.'' They think that having a ``clear, operational definition of AGI'' is the way to do this.

At first, their goal seems to be purely deflationary---they want some shared, operationalizable terminology so that we know when we are referring to similar phenomena. \citet{morris_levels_2023} are inspired by the Levels of Automation framework \cite{sae_international_j3016_202104_2021} used to grade self-driving car technology---a framework that was also allegedly adopted to enhance communication.\footnote{However, see Section \ref{fake-deflationary} for a discussion of how this framework also drifted from its initial deflationary goals.} Their deflationary approach is further underlined by a willingness to change the definition of AGI in the future if we gain the ability to operationalize things that are currently not operationalizable. For example, they currently want to define AGI in terms of ``capabilities rather than processes'' because we currently have little insight into the underlying mechanisms of many AI systems. They allow that research into mechanistic interpretability might eventually lead to operationalizable definitions involving processes, which then ``may be relevant to future definitions of AGI'' \cite{morris_levels_2023}. 

Another AGI account that is on the deflationary spectrum is that of \citet{adams_mapping_2012}, who have ``a pragmatic goal for measuring progress toward its attainment''. The authors acknowledge that ``The heterogeneity of general intelligence in humans makes it practically impossible to develop a comprehensive, fine-grained measurement system for AGI''. However, in order to measure progress, they would still like ``a common framework for collaboration and comparison of results''. This conception is similar to that of \cite{morris_levels_2023} in treating the AGI framework as a common language for communication purposes. To the extent that \citet{adams_mapping_2012} adopt requirements for ``general cognitive architectures'', they emphasize that these requirements are not final and are ``simply \dots a convenient point of departure for discussion and collaboration''.

We'll see in Section \ref{fake-deflationary} that these deflationary accounts depart from their initial declared intentions and end up including some value-laden choices. But first, we'll take a tour of some non-deflationary accounts of AGI.

\subsection{Value-laden choices in AGI definitions: a taxonomy}
\label{dimensions-taxonomy}

Having outlined deflationary views on AGI, we now identify some value-laden choices that non-deflationary views make in their conception of AGI or human-level AI. We organize these views by the dimensions on which these choices are made (see Appendix \ref{table} for a summary of these dimensions with examples).  Many of these choices are better understood as differences in focus and emphasis, rather than as mutually exclusive assumptions about the correct conception of AGI.

\subsubsection{Economic value}
\label{economic}
Some definitions of AGI are explicit about the values that they are centering. For example, OpenAI's charter claims that ``OpenAI's mission is to ensure that artificial general intelligence (AGI)---by which we mean highly autonomous systems that outperform humans at most economically valuable work---benefits all of humanity'' \cite{openai_2018}. This definition of AGI is meant to track the metric of \emph{performing economically valuable work}. As others have pointed out \cite{morris_levels_2023}, this assumes that other types of work are less valuable---a normative assumption.

OpenAI’s definition is perhaps the most explicit about its values. But we can find other definitions that embed similar values. \citet{suleyman_coming_2023}, for example, define ``Artificial Capable Intelligence'' by its ability to pass what they call the ``Modern Turing Test'': the capability to turn a starting pot of \$100,000 of capital into \$1,000,000 over several months. Similarly, \citet{nilsson_human-level_2005} proposes an ``employment test'' to replace the Turing Test: ``To pass the employment test, AI programs must be able to perform the jobs ordinarily performed by humans. Progress toward human-level AI could then be measured by the fraction of these jobs that can be acceptably performed by machines.''

In contrast, \citet{morris_levels_2023} reject definitions based on economic value alone. They think that benchmark tasks should align with what people value beyond economic value, including metrics that are harder to automate or quantify. Here, in tension with their purported deflationary orientation (see Section \ref{deflate}), they explicitly argue for having components of AGI that are less operationalizable (because they are harder to quantify). The reason they are doing so is value-laden: because they think that there are values beyond economic value that are worthwhile to aim for.

\subsubsection{Embodiment}

The choice of whether to include the ability to carry out tasks in the physical world is another dimension on which AGI frameworks differ. \citet{morris_levels_2023} make an unexplained choice to exclude embodied tasks from their framework. This exclusion is particularly mysterious because physical tasks are, on the face of it, no harder to operationalize than non-physical tasks.

In contrast, criteria like Wozniak's coffee test require embodied AI systems. This test, first proposed by Apple founder Steve Wozniak \cite{coffee}, tasks the machine with going into an ``average'' American home and making coffee in it. It has since been included in definitions of AGI \cite{goertzel2012architecture, Marcus_2022}. AI critics like \citet{weizenbaum_computer_1976} have also cited the field's relative inattention to embodiment as a major shortcoming.

The decision to include or exclude embodiment in the definition of AGI is value-laden, as it incorporates assumptions about which tasks are considered valuable. Privileging the mental over the physical is a normative choice. It also carries normative consequences, such as influencing the types of AI systems likely to be developed and their potential impacts on society.

\subsubsection{Human-like processes versus outcomes} 
\label{anthropomorphic}

Another divisive choice in conceptualizing AGI or human-level AI is whether to insist on mechanisms or processes that mimic human cognitive processes. For example, \citet{morris_levels_2023} want to prioritize outcomes (``what an AGI can accomplish'') over human-like processes because outcomes are more operationalizable. Excluding ``processes'' from the definition means that criteria based on the following are excluded: consciousness \cite{butlin_consciousness_2023, lenharo_ai_2024, searle_minds_1980, summerfield_natural_2023, smart_beyond_2015}, sentience \cite{schwitzgebel_defense_2015}, and abilities modeled on human children \cite{turing_computing_1950, nilsson_human-level_2005, gopnik_ais_2019, gopnik_what_nodate, summerfield_natural_2023}.

This choice of excluding anthropomorphic cognitive processes goes beyond purely conceptual questions about how to best define AGI.\footnote{Here is a possible epistemic argument for taking a stance on the exclusion of anthropomorphic processes from the concept of AGI. The question of whether or not implementing human-like processes is needed to achieve human-like outcomes is an empirical question. Settling open-ended empirical questions through definitions is bad epistemic practice. Definitions of AGI should, therefore, be agnostic on empirical questions such as the importance of human-like processes to achieve different human-like outcomes. We see this as amounting to providing a principled reason for \emph{not answering} the question of whether AGI needs to involve anthropomorphic cognitive processes. Discourse on AGI often goes beyond this agnostic stance for practical reasons.} It also influences the types of research projects that attract AI investment, which has ethical consequences in terms of what the AI we build (and its attendant social consequences) looks like. \citet{lenharo_ai_2024} interviews a researcher claiming that ``to his knowledge, there was not a single grant offer in 2023 to study the topic'' of consciousness in AI.

Rich Sutton's ``The Bitter Lesson'' articulates this practical concern about maximizing return on investment \cite{sutton_bitter_2019}. Sutton frames this as a choice between focusing on implementing human-like processes versus focusing on ``general methods that leverage computation''. He frames the choices as practically exclusive in the sense that ``[t]ime spent on one is time not spent on the other'' \citet{sutton_bitter_2019}. The bitter lesson is that building human-like processes into AI may be psychologically satisfying to researchers, but tends to be much less effective than focusing on ``general purpose method[s]'' that scale with computation. 

However, it's not clear that the practical argument for ignoring the processes underlying intelligence is \emph{correct}. \citet{summerfield_natural_2023}, for instance, makes the case that in drawing lessons from ``natural general intelligence'' (i.e. human cognition and intelligence), work on AGI stands to yield more ``useful'' and ``effective'' systems, and to better leverage the ``tight intellectual synergy between AI research, cognitive science, and neuroscience''.

In short, different practitioners make different choices about what they consider to be ``practical'' for achieving AGI. Choices like these about effectiveness and how to best allocate limited resources are inherently value-laden---raising questions like \emph{for what goals, given whose beliefs and preferences, given what conditions}? For example, \citet{summerfield_natural_2023} does not articulate what he means by ``useful'' or ``effective'', let alone for whom it is useful or effective. Similarly, when \citet{morris_levels_2023} argue for having shared, operationalizable (and thus outcome-based) standards for AGI on the grounds that the standards will be ``useful'', they do not articulate whom it will be useful for, or consider whether it may be less than useful to some. Critics of current trends in AI have pointed to how its harms and benefits are distributed unevenly in ways that are correlated with existing power structures \cite{green_data_2021, abebe_roles_2020, blodgett_language_2020, birhane_towards_2021, costanza-chock_design_2020}. Given this context about how AI is currently used, implicit assumptions that ``useful'' means ``useful for everyone'' need more justification. 

\subsubsection{Generality}
\label{generality}

A point of widespread agreement among accounts that specifically favor the term ``AGI'' over ``human-level intelligence'' is the importance of \emph{generality}  \cite{goertzel_artificial_2014, morris_levels_2023, aguera_y_arcas_artificial_2023, summerfield_natural_2023}. \citet{summerfield_natural_2023} summarizes this as the view that a ``generally intelligent individual is polymathic---good at  everything.''\footnote{\citet{summerfield_natural_2023} emphasizes the importance of the kind of flexibility that humans display in ordinary everyday activities---``whether performing daily rituals of basic hygiene, navigating the streets of their neighbourhood, negotiating the local market, deftly parenting their children, or judiciously managing household finances''---rather than in the kinds of abilities associated with ``chess grandmasters'' or ``Nobel laureates''. This sense of generality is quite different from what gets emphasized by \emph{g}. Rather than centering a putative variable that correlates with \emph{improved performance} across all cognitive tasks, Summerfield emphasizes the multiple complex facets of the kind of cognitive flexibility that humans display.} The recent popularization of the term ``AGI'' is often traced to the mid-2000s interest in contrasting ``narrow AI'' capable of ``speciﬁc `intelligent' behaviors in speciﬁc contexts'' with systems that would be able to ``self-adapt to changes in their goals or circumstances'', ``generalize knowledge from one goal or context to others'', and so on \cite{goertzel_artificial_2014}.\footnote{\citet{goertzel_artificial_2014} traces the genealogy of the term ``AGI'' as follows. ``The brief history of the term ``Artiﬁcial General Intelligence'' is as follows. In 2002, Cassio Pennachin and I were editing a book on approaches to powerful AI, with broad capabilities at the human level and beyond, and we were struggling for a title. I emailed a number of colleagues asking for suggestions. My former colleague Shane Legg came up with ``Artificial General Intelligence,'' which Cassio and I liked, and adopted for the title of our edited book [\citet{goertzel_pennachin_artificial_2007}]. The term began to spread further when it was used in the context of the AGI conference series. A few years later, someone brought to my attention that a researcher named Mark Gubrud had used the term in a 1997 article on the future of technology and associated risks [\citet{gubrud_nanotechnology_1997}].''}

Although widespread, the strong emphasis on ``generality, adaptability and ﬂexibility'' \cite{goertzel_artificial_2014} in machine intelligence is not universal among accounts of human-level machine intelligence. For instance, \citet{bostrom_superintelligence_2014} departs from this norm by denying that achieving human-level intelligence or superintelligence requires ``general'' intelligence at all. Bostrom sees this as part of his strong rejection of the requirement for human-like cognitive processes. He instead remains agnostic about the structure of abilities needed to match or outclass the instrumental performance of human intelligence.

We argued earlier (Section  \ref{anthropomorphic}) that assumptions about whether AGI should include human-like processes or not is value-laden because it is based on notions of ``usefulness'' or ``effectiveness'' that presuppose whom and what the assumptions are useful for. The same line of thought applies to arguments about whether the generality of abilities is important. Assumptions about what practical outcomes are desirable are needed to make the case for or against the importance of generality. Bostrom's line of argument presumes that replicating the \emph{instrumental} performance of human intelligence is the goal---but there are other possible goals for AI. Similarly, others' assumption that being polymathic is essential to AGI glosses over two points: whether this is a desirable goal for AI, and the real diversity of human abilities (not all humans are similarly polymathic). Here we see an analogy with critiques of \emph{g} on the basis that it ignores human diversity (see Section \ref{ableism}).

\subsubsection{Individualism}
\label{individualism}

Conceptions of AGI also differ on whether they conceive of intelligence as a property of individuals---such as isolated humans or systems. One aspect of individualism about AGI is \emph{ontological}, having to do with whether the entity ascribed ``intelligence'' is an individual. The psychometrics project of measuring and comparing the intelligence and cognitive skills of humans takes individuals as a key unit of analysis. AI evaluation practices like benchmarking, as currently practiced, mostly treat individual models as bearers of the properties they measure, and definers of AGI often propose tests that are mostly or entirely tests on individual agents \cite{chollet_measure_2019, morris_levels_2023}.

By contrast, accounts like \citet{bostrom_superintelligence_2014} and \citet{attard-frost_queering_2023} reject ontological individualism. Bostrom argues that when thinking about what it means for machines to match or outclass human intelligence, the relevant unit of comparison includes not just the intelligence of an individual human but also ``the combined intellectual capability of all of humanity'' at present \cite{bostrom_superintelligence_2014}. Specifically, he argues that collectives---such as ``firms, work teas, gossip networks, advocacy groups, academic communities, countries, even humankind as a whole'' \cite{bostrom_superintelligence_2014}---can be understood as a mechanism for increasing intelligence. Bostrom also refers to collective intelligence as ``collective intellectual problem-solving capacity''. This rejection of ontological individualism is related to the longer tradition of thinking about collective intelligence \cite{levy_lintelligence_1994, levy_social_2010, engelbart_dc_augmenting_1962, baltzersen_cultural-historical_2021, suran_frameworks_2021, araya_augmented_2023, landemore_collective_2012, anderson_epistemology_2006, putnam_reconsideration_1989, dewey_creative_1939, peters_toward_2015}. \citet{baltzersen_cultural-historical_2021} proposes thinking of collective intelligence as ``collective problem solving'', both in large and small groups of people.\footnote{\citet{levy_social_2010} similarly defines collective intelligence as ``the capacity of human collectives to engage in intellectual cooperation in order to create, innovate and invent''.} In Section \ref{democracy}, we come back to a strand of this tradition that frames \emph{democracy} as a ``precondition for the full application of intelligence to the solution of social problems'' \cite{putnam_reconsideration_1989}.

Ontological individualism has normative implications in shaping the ``north star'' goals of AI: is the imagined challenge building machines that rival the intelligence of isolated human individuals, or creating systems that replicate the collective intelligence of groups or entire societies? It is unclear, on the face of it, which goal is ethically preferable to the other (of course, not building either is also an option). Due to contemporary AI's focus on individual intelligence, it is hard to envision how AI systems would be different if the focus was instead on collective intelligence---but it is plausible that that counterfactual would lead to very different AI systems, with correspondingly different impacts on society.

A second dimension of individualism in conceptions of AGI is methodological: concerning whether measurements of intelligence are carried out in environments where individuals are acting by themselves or where agents interact in a social environment with other agents.  Most AI benchmarks are focused on individual agents carrying out tasks by themselves. But an alternative vision is possible \cite{attard-frost_queering_2023}---one that we describe in more detail in Section \ref{blair}. These two different ways of measuring intelligence would lead to different prioritizations of resources for research and development, with attendant ethical implications. For example, if we measure intelligent agents' performance in more social and interactive environments, would this also lead to AI systems having fewer unanticipated harmful effects when deployed into the ``real world''? Arguably, many of the recent cases of unanticipated harmful effects from AI systems being deployed are due to these systems being evaluated pre-deployment on decontextualized tasks that ignore aspects of the social environment \cite{wolf2017we, ganguli, saisubramanian2022avoiding}.

\subsubsection{Instrumentality Thesis} 
\label{instrumental}

A widespread assumption in discussions of AGI is what \citet{russell_human_2019} calls the ``standard model of intelligence'': ``[Humans or] machines are intelligent to the extent that their actions can be expected to achieve their objectives.'' Similarly, \citet{legg_collection_2007} characterize intelligence as ``measur[ing] an agent’s ability to achieve goals in a wide range of environments.'' These accounts endorse what we call the \emph{instrumentality thesis}: the view that intelligence is a means to whatever ends, final goals, or preferences an agent might have.\footnote{\citet{bostrom_superintelligence_2014} calls this the orthogonality thesis, which he defines as follows \cite{bostrom_superintelligence_2014}: ``Intelligence and final goals are orthogonal: more or less any level of intelligence could in principle be combined with more or less any final goal.''  We find it more helpful to highlight the similarity of this conception of intelligence with the conceptions of instrumental rationality that come from traditions like utility theory, or embrace Hume's adage that ``Reason is, and ought only to be the slave of the passions'' \cite{hume_treatise_1991, sep-emotions-17th18th}.} These accounts conceive of being more or less intelligent as categorically different from being better or worse at determining what matters, what goals are worth pursuing, or what is good for its own sake. \citet{bostrom_ethical_2003} illustrates this feature of intelligence with the example of a paperclip maximizer: an agent with the final goal of building a world with maximally many paperclips. Paperclips are sometimes useful tools, such as for organizing paper documents. But paperclip maximizing as a final goal is patently not worthwhile. The instrumentality thesis is part of a common tradition in AI: thinking of intelligent agents as maximizing the expected utility of actions for satisfying given objectives or preferences.

The assumption that intelligence is purely about instrumental rationality is another juncture where values come into play. An alternate conception of intelligence could consider the ability to determine what goals are worth pursuing as a form of intelligence. Agents with this ability would be able to form independent views about whether the goals they are assigned by other agents are worthwhile, and if not, how or whether to resist their assignments. In short, the view that intelligence encompasses only instrumental rationality lays the ground for AI to develop agents that can optimize the goals of their designers, but that have no ability to question those goals. We can imagine instead a very different future of AI: where we have intelligent agents that are able to dissent about final goals (see Section \ref{democracy}). Which future is more desirable is a question of values: do we want intelligent machines that are completely subservient, or do we want machines that can collaborate with humans in determining what goals are worth pursuing?

\subsubsection{Operationalizability/Measurability}

Another dimension on which accounts of AGI differ is whether to require operationalizability or measurability for their definition. At one end of the spectrum, we have \citet{morris_levels_2023}, who frame their project around operationalizability and explicitly reject criteria that are not operationalizable. However, other discussants include definitions that are not measurable. \citet{bostrom_superintelligence_2014}, for example, makes frequent reference to the collective intelligence of all of humanity at present in his framework---a concept that seems difficult, if not impossible, to operationalize.  

Operationalizability may appear value-neutral at first glance. However, scholars have argued that simply having a common language and shared standards is in itself value-laden. Commensuration (the act of establishing common standards to measure things) allows large institutions to coordinate actions and automate decision-making, but less powerful groups often argue that some things cannot be measured in order to resist actions that can be justified through ``rational choices'' \cite{espeland1998commensuration}. Social scientists have also recognized that commensuration has performative effects, creating a kind of path dependency: ``as commensuration gets built into practical organizations of labor and resources, it becomes more taken for granted and more constitutive of what it measures'' \cite{espeland1998commensuration}. This type of path dependency has been recognized in various domains of commensuration, such as standardized grades of grain quality \cite{porter1996trust} and university rankings \cite{fowles2016university}. Previous work on ML benchmarking has also highlighted how benchmarking creates performativity and path dependency \cite{dehghani_benchmark_2021,blili-hamelin_borhane_making_2023}.

\subsubsection{Choice of tasks/benchmarks}

One cross-cutting dimension of difference across definitions of AGI is the tasks or benchmarks that they include in their criteria for AGI. Some of the previously mentioned dimensions intersect with this---for example, if embodiment is part of your AGI definition, you would include some embodied tasks in your definition. In contemporary AI research, with its plethora of tasks and benchmarks, researchers sometimes make choices about which tasks they emphasize when estimating progress toward AGI.

Two prominent AI researchers recently argued that ``[t]he most important parts of AGI have already been achieved by the current generation of advanced AI large language models'' \cite{aguera_y_arcas_artificial_2023}. While these researchers define AGI roughly as instructable systems able to operate over a wide variety of topics, tasks, modalities and languages, the fact that the strengths of current LLMs are considered to already cover ``the most important parts of AGI'' implies that the things that LLMs cannot do are the less important parts of AGI. These LLM weaknesses include (as they acknowledge themselves) arithmetic and adhering to facts. However, they do not explain why these areas of reasoning are devalued relative to LLM strengths like translation or synthesizing code. Other researchers, in contrast, believe that arithmetical and logical reasoning are necessary components of AGI \cite{Marcus_2023}. \citet{aguera_y_arcas_artificial_2023}'s assertion that the most important aspects of AGI are achieved by current-day LLMs is value-laden. Choices of tasks and benchmarks are a key site of value-laden design choices \cite{blili-hamelin_borhane_making_2023, bommasani_evaluation_2022}. Deciding to focus on LLMs over deductive reasoning systems has ethical implications because these systems will change our societies in different ways. \footnote{For another critique of the reliance on benchmarks in relation to AGI, see \citet{summerfield_natural_2023}.}

\subsubsection{Importing \emph{g} into AI}
\label{g-ai}

One family of AGI characterizations relies on directly importing the concept of \emph{g} from human intelligence into AI. These papers assume that the hypothesis that \emph{g} is an explanatory causal factor in human intelligence is true, then seek to discover it in AI systems. For example, \citet{hernandez-orallo_general_2021} try to find \emph{g} by conducting factor analysis on the results of machine experiments. They motivate this by analogy to the supposedly explanatory role of \emph{g} in human intelligence, which they do not question. This move has the following weaknesses.\footnote{For another critique, see also \citet{russell_human_2019}.}

Firstly, the critiques of factor analysis as a methodology for finding a unique causal structure, as applied to the case of \emph{g} in human intelligence (see Section \ref{factor-analysis}), would also transfer over to the case of AI. Secondly, taking AI to have a directly analogous \emph{g}  factor means that all the value-laden assumptions around what human abilities count as important are imported over into the AGI case (see Section \ref{human}). For example, \citet{chollet_measure_2019}, who also draws inspiration from \emph{g}, takes for granted that there is a ``space of tasks and domains that fit within the human experience'' that we should use to measure intelligence---ignoring critiques of similar assumptions in the human intelligence literature (see Section \ref{ableism}).

\subsection{Are deflationary views truly deflationary?}
\label{fake-deflationary}

We've outlined dimensions on which accounts of AGI make value-laden choices that aren't determined by purely epistemic criteria. We also saw that design choices made within the more ``deflationary'' views of AGI  go beyond the apparently value-neutral goal of having a common language and shared standards. Here we discuss some more subtle ways in which these ``deflationary'' views incorporate values.

\citet{morris_levels_2023} propose a common language in order to enable us to compare models, assess risks, and measure progress. These are different goals that a common language can potentially achieve. But it is not clear that each goal would be maximized by the \emph{same} language. Should the shared standards that would be ``best'' for risk assessment necessarily be the same as those that would be best for measuring progress? To determine what shared standards would be best for risk assessment, we need to make some normative assumptions---for example, about which risks are more important. Our shared standards might differ depending on whether we think so-called ``existential risk'' is the biggest risk, in contrast to ongoing harms from AI.\footnote{For commentary on how policymakers are over-focused on existential risk, see \cite{hanna_bender} and \cite{gomez}.} As an example, later on in the paper, \citet{morris_levels_2023} reject \citet{suleyman_coming_2023}'s definition of AGI (see Section \ref{economic}) because it might introduce ``alignment risks''. It follows that minimizing alignment risks is at least an implicit desideratum for \citet{morris_levels_2023}'s framework---but this is not stated upfront.

Looking at another deflationary view of AGI, \citet{adams_mapping_2012}'s way of framing the quest for AGI, as described in Section \ref{deflate}, assumes that their ``pragmatic goal'' is worth achieving---a topic on which people with different values might disagree.

Furthermore, there are insights to be drawn from \citet{morris_levels_2023}'s analogy to the Society of Automotive Engineers' (SAE) framework of levels of automation for automobiles \cite{sae_international_j3016_202104_2021}. The intention of the analogy was to justify a similar move in the AI space, with the implied utility of a common language and shared standards. However, critical work on the SAE framework has argued that what appears to be merely a descriptive, technical definition with the SAE's stated intentions of ``simplifying communication'' and ``providing clarity'' in fact has normative assumptions and implications \cite{hopkins2021talking}. Implications teased out by \citeauthor{hopkins2021talking} include: promoting ``homogeneity in mobility futures'', specifically a homogeneous vision of automation as the future of mobility; reproducing a ``dominant discourse of an expert-led, technologically-centred vision of mobility futures''; and removing ``obstacles and impediments to the successful (and timely) development of the automated vehicle niche.'' Similarly, we may ask if shared standards for AGI promote homogeneity in AI futures and reproduce an expert-led dominant discourse. Certainly, the standards currently being developed are expert-led and lack input from people who are more likely to suffer negative impacts in AI---an issue that we discuss further in Section \ref{epistemic-justice}.

\section{Towards contextualized, politically legitimate, and social intelligence}
\label{towards}
We now outline alternative visions for values worth centering in imagining future forms of machine intelligence. These views embrace the value-laden nature of intelligence instead of side-stepping it. Broadly speaking, the following proposals take seriously the role of physical and social contexts in definitions of machine intelligence, in contrast to the views outlined in Section \ref{choices}. We selected these views because they embody the values of contextualism, epistemic justice, inclusiveness, and democracy, which we consider vital for visions of the future of AI worth pursuing.

\subsection{Contextual intelligence}
\label{blair}

The importance of a contextual understanding of sociotechnical systems is well-recognized \cite{selbst_fairness_2019, weidinger_sociotechnical_2023, nist_ai_2023, lazar_ai_2023, shelby_sociotechnical_2023, mohamed_decolonial_2020}. Given the interlocking social and technical factors that shape the impact of AI systems, we need to consider ``how a system is used, its interactions with other AI systems, who operates it, and the social context in which it is deployed'' \cite{nist_ai_2023}. \citet{attard-frost_queering_2023}’s account of intelligence as ``value-laden cognitive performance'' brings a contextualist perspective to conceptualizing intelligence itself. 

Their account centers the role of values in defining what counts as good cognitive performance, in addition to the role of ``interdependencies between agents, their environments, and their measurers in collectively constructing and measuring context-specific performances of intelligent action'' \cite{attard-frost_queering_2023}.  As they argue, contextually situated accounts of intelligence have a history in STS \cite{hayles_unthought_2017, barad_posthumanist_2003}, but also in AI \cite{weizenbaum_computer_1976}.

Attard-Frost’s account of intelligence contrasts with more individualistic views that measure intelligence ``with reference to an extremely constrained and highly standardized set of cognitive activities performed by individuals, rather than with reference to situated activities performed in relation to other individuals and social environments'' \cite{attard-frost_queering_2023}. To use their lab analogy, Attard-Frost’s account is \emph{in vivo} rather than \emph{in vitro}---an account of how cognition is performed in social environments.\footnote{See also \citet{hutchins_cognition_1995} on the distinction between cognition ``in the wild'' and cognition under artificially controlled (laboratory) conditions.} Another way of viewing the difference is that \emph{in vivo} accounts have higher ecological validity---a property that even the less contextual accounts of AGI agree is desirable \cite{morris_levels_2023}.

Attard-Frost’s account may have some practical drawbacks---perhaps it is harder to operationalize, or to compare systems that operate in different contexts. Thus, whether or not we choose to adopt something like their account is itself dependent on our values and goals: their goal of queering intelligence is itself a normative one, and it may conflict with other normative goals.

\subsection{Inductive risk and social values}
\label{inductive}

Another helpful lens on how to define future forms of machine intelligence is that of inductive risk. In the values in science literature, inductive risk is often used to justify having different standards of evidence for accepting scientific claims. The idea is that when the potential social costs of accepting/denying a scientific claim are high, social and political values \emph{should} influence standards of evidence \cite{douglas2000inductive, steel2010epistemic, blili-hamelin_borhane_making_2023}. 

We can adapt the argument from inductive risk to the case of defining AGI by reframing it in terms of standards of evidence for accepting definition(s) of AGI.\footnote{We frame it this way because it is unclear whether definitions of AGI are ``scientific claims'' as traditionally construed by philosophers of science. We thank Ravit Dotan for proposing this point.} As Section \ref{choices} has illustrated, current AGI definitions make many choices that are unjustified by epistemic values, or are based on implicit assumptions about what is socially valuable. Adapting the argument from inductive risk to AGI, we would say: due to the significant social impacts that could arise from having definitions of intelligence that value certain social goals/tasks/beings over others, social and political values should influence how we define AGI.

\subsection{Epistemic justice for defining future forms of AI}
\label{epistemic-justice}

One aspect of many definitions of AGI that we have analyzed here is that they mostly come from actors who have relatively more power and influence over the future of AI. At the same time, there is a dearth of voices in the conversation from people who are more likely to be harmed by deployed AI systems. Advocates of AGI and human-level AI imagine this technology as impacting almost everyone. Taking that ambition seriously requires processes that give a meaningful say to the communities who would be impacted by the technology. Dreams of future technologies should come hand in hand with participatory, inclusive, and---as we argue below--- politically legitimate decision-making processes.
  
A positive vision for defining future forms of AI (or, taking a step back, even deciding if it should be pursued) would learn from the participatory ML and epistemic justice literature. We would love to see a vision for future forms of machine intelligence that is constructed through participatory methods \cite{birhane_participatory, delgado_participatory, young_participation_2024}, while still being aware that the products of these methods have their limitations \cite{sloane_participation_2020}. Needless to say, these participatory methods should take care to include the perspectives of not just those who design or fund AI, but those who will be impacted by it in other ways.

The epistemic justice lens also highlights discussions of epistemic justice in scientific fields that have had to make similar assessments of difficult-to-measure concepts. Epistemic justice is the idea that different groups of people have, due to power differentials, differing levels of credibility or differing contributions to the concepts that underpin our shared knowledge, and this is an injustice because it can render unequal distributions of consequences (e.g. misunderstanding what sexual harassment is has more impact on women than on men) or make it harder for less powerful groups to understand their own experiences \cite{fricker_epistemic_2007, schmidt_epistemic_2019}.

\citet{alexandrova_democratising_2022} argue that scientific fields studying thick concepts like intelligence, health, wellbeing, and sustainability have a distinctive need for participatory processes. For research that has implications on real-world communities---such as fields influencing policy and lawmaking---they argue that relying on the personal values of people studying the phenomena is not enough. Research on thick concepts that affect real-world communities, they argue, ought to seek \emph{legitimacy} for its values. Political legitimacy concerns how states and institutions avoid coercion and oppression in their use of power over their own citizens, such as through justifying their power to those they have power over \cite{peter-sep-legitimacy}. \citet{alexandrova_democratising_2022} propose an epistemic constraint on those fields, requiring the values of the field to be legitimated through a participatory ``political process that includes all the stakeholders of this research.'' 

Similarly, \citet{elabbarcuratorial} has extensively examined how values enter into Intergovernmental Panel on Climate Change (IPCC) reports. He considers these reports to involve difficult curatorial choices about ``which truth-apt claims and representations to display in a given space at all, under rationing pressure''. These curatorial choices are value-laden. For example, 89\% of government reviews in the IPCC's Summary for Policymakers are from developed nations, indicating one way that powerful nations have more input into the report. Crucially, the climate change case points to the ineffectiveness of interventions like transparency in design choices. Elabbar argues that being transparent about the role of values in justifying design choices fails to empower many stakeholders. This is because ``developing emission figures presumes advanced knowledge of carbon accounting methods and various other forms of specialist expertise'', and ``[i]n cases where the effect of value choices is subtle and complex, lay audiences will typically lack the capacity to cash out transparency in the form of genuine alternatives that accord with their values''. Thus, he rejects transparency as a solution to this type of epistemic injustice.

We think a similar situation applies to the problem of defining AGI. AGI definitions are currently being developed by people who, as technical experts designing AI systems, have relatively more power, and the effects of their design choices are complex. Transparency into this process would likely be insufficient to empower most stakeholders to imagine alternatives to these visions---which is not to say that it isn't desirable. Participatory methods are thus a key part of the solution.

Notably, by suggesting more participatory methods, we do not mean that every issue surrounding future forms of machine intelligence should be subject to a population-wide referendum. As the IPCC case illustrates, specialist expertise is indispensable to making good decisions on these issues. Luckily, modern democracies have several methods of ensuring that citizens have a voice in state decisions through mechanisms other than referenda. For example, representative democracy, where citizens elect representatives who make decisions on behalf of them, mixes elements of expert-led decision making with citizen participation. Similarly, polities have also experimented with citizen assemblies on important issues \cite{ireland_citizens, huang2017public}. These are small forums where a selected set of ``ordinary'' citizens debate issues as a mechanism for influencing public opinion and political decision-makers.

\section{Conclusion: politically legitimate intelligence}
\label{democracy}
Donna Haraway's ``A Manifesto for Cyborgs'' argues that despite its origin in ``racist, male-dominant capitalism'' and ``militarism'', the figure of the human-machine hybrid can be repurposed towards resisting oppressive social categories in imagining the future \cite{haraway_manifesto_1987, forlano_cyborg_2024}. Imagining ``altering bodily functions'' to allow humans to survive in space \cite{clynes_cyborgs_1960} opens the door to Ashley Shew's interrogation of how technology reinforces ableism:``Everyone in space will be disabled'' \cite{shew_disabled_2018, forlano_cyborg_2024}. The scale, power, and scope---or perhaps lack of scope, as Gebru and Torres argue \cite{Gebru_Torres_2024, burrell_introduction_2024}---of dreams of AGI should raise serious doubt about whether the concept is worth similarly subverting. Whether through subverting or discarding AGI, we nevertheless believe in the need for more contextual, participatory, and democratic approaches to imagining future forms of intelligence and machines.

To that end, we have taxonomized the discordant value-laden choices of AGI discourse. We surfaced alternative paths that provide more contextual, participatory, and epistemically just perspectives on imagining future forms of AI. We suggested that the question of what forms of machine intelligence are worth pursuing calls for a plurality of contested and value-laden perspectives. We now conclude by highlighting perspectives that provide positive rejoinders for placing dissent, deliberation, and political legitimacy at the center of conversations about intelligence and future technologies. 

Democratic legitimacy and social conceptions of intelligence are connected. \citet{birhane_participatory} defend the need for participatory AI to be supplemented with approaches that yield ``stronger forms of validation and legitimacy'' through democratic governance. Though they do not refer to it, their position is resonant with an intellectual tradition that sees democratic institutions as embodying a distinctive form of \emph{intelligence} \cite{anderson_epistemology_2006, anderson-dewey-moral, putnam_reconsideration_1989, landemore_democratic_2013, festenstein_does_2019, alexander_educating_2021, dewey_creative_1939, festenstein_deweys_2023, anderson_john_1991, mill_liberty_2003}. This tradition asks the question: how can we arrive at empirically informed solutions to social problems that avoid subordinating people to others? The answer is to consider democracy as its own form of ``social intelligence''---as a collective process of deliberation and reasoning that can lead to iterative improvement on solutions and worthwhile goals.

Building on John Dewey's conception of ``democracy as the use of social intelligence to solve problems of practical interest'', \citet{anderson_epistemology_2006} argues that democratic institutions embody solving social problems without resorting to oppression or coercion. She argues that democracies are uniquely suited to investigating solutions to \emph{public interest} problems, problems that need to be solved through deliberation (``votes and talk'') rather than through procedures like markets. They do so by pairing sources of legitimacy like procedural fairness and universal inclusion (through mechanisms like law, rights, and voting), experimentation (``revising [\dots] decisions on the basis of experience with their consequence''), and dissent \cite{anderson_epistemology_2006}.

Public interest solutions to problems like deforestation and sustainability \cite{agarwal_participatory_2001} require institutions capable of correcting their shortcomings and unintended consequences.\footnote{Anderson illustrates these features through \citet{agarwal_participatory_2001}'s study of the effects of gender exclusion in limiting the equity and the efficiency of community forestry groups (CFG) as sustainable solutions to deforestation in India and Nepal. Agarwal highlights a gendered division of labor that made women the primary users of local forests while imposing ``formal and informal obstacles'' to their participation in CFGs \citet{anderson_epistemology_2006}---such as the combination of rules that limit participation to 1 per household with patriarchal norms that favor men as representatives of households, and ``CFG meeting times that coincide with women’s household tasks''.} ``Just as the solution to scientific problems is to do more science, the cure for the ailments of democracy is more democracy'' \cite{anderson_epistemology_2006}.  Similarly, for Dewey, this process of appraisal, discussion, and judgment is what constitutes ``organized intelligence'':
\begin{quote}
    Of course, there are conflicting interests; otherwise there would be no social problems [\dots] The method of democracy---inasfar as it is that of organized intelligence---is to bring these conflicts out into the open where their special claims can be seen and appraised, where they can be discussed and judged in the light of more inclusive interests than are represented by either of them separately. \cite{dewey_liberalism_1987}\footnote{Quoted by \citet{festenstein_does_2019}.}    
\end{quote}

\citet{putnam_reconsideration_1989} summarizes this view: democracy ``is the precondition for the full application of intelligence to the solution of social problems''.\footnote{See \citet{festenstein_does_2019} for an overview of critiques of this family of views, both as interpretations of Dewey and as conceptions of democracy.} This lens puts pressure on the narrow range of problems, tasks, and processes that take center stage in discussions of would-be human-level AI and AGI. Intelligent solutions to social problems should not be framed as a matter of finding optimal means for satisfying fixed preferences. Rather, they require procedures that support universal inclusion in interrogation, deliberation, and dissent about what counts as the ``common good'' \cite{putnam_reconsideration_1989} and the ``public interest'' \cite{anderson_epistemology_2006}. They require procedures with political legitimacy \cite{peter-sep-legitimacy}.  On this view,  future forms of AI can provide intelligent solutions to social problems only if they are also objects of dissent and deliberative contestation, rather than systems designed from the top-down by ``experts'' only. The project of imagining ``worlds'' \cite{costanza-chock_design_2020} and future technologies worth building should be one of collective ``experiments in living'' \cite{mill_liberty_2003, anderson_john_1991}, in which all impacted rights-holders hold decision-makers accountable.

\section*{Acknowledgments}

We are thankful to Ravit Dotan for extensive input on early versions of this paper (see also \ref{inductive}). This paper took shape in response to a practical problem we faced in an ongoing position paper project on AGI: the lack of homogeneity in conceptions of AGI. We felt that a paper dedicated to investigating definitions of AGI would help move the broader project forward. We are deeply grateful to our collaborators on the ongoing position paper project: (in alphabetical order) Chris Graziul, El Mahdi El Mhamdi, Hananel Hazan, Katherine Heller, Mariame Tighanimine, Margaret Mitchell, Olivia Guest, Roxana Daneshjou, Shiri Dori-Hacohen, Talia Ringer, and Todd Snider.

\bibliography{agi}

\appendix

\section{Dimensions of AGI: A Summary}
\label{table}

We have described various accounts of AGI and highlighted choices these accounts make that are at least partially value-laden. The following table summarizes some of the dimensions on which various accounts of AGI differ, highlighting how each account is making a choice about how to define intelligence. See Section \ref{dimensions-taxonomy} for more details on each dimension.

\begin{landscape}

\begin{longtblr}[
  caption = Summary table of how different conceptions of AGI differ on value-laden dimensions,
  label = dimensions-table,
  entry = none,
]{
  width = \linewidth,
  colspec = {Q[100]Q[347]Q[347]},
  hlines,
  vlines
}
\textbf{Dimension}                                   & \textbf{Description}                                                                                                                                                                                                                                                                                                           & \textbf{Examples}                                                                                                                                                                                                                                                                                                                                                                                                                                 \\
Embodiment                                  & Whether the ability to accomplish certain physical tasks is part of the definition                                                                                                                                                                                                                                    & {Required: \cite{coffee,Marcus_2022,weizenbaum_computer_1976} \\Not required: \cite{morris_levels_2023}}                                                                                                                                                                                                                           \\
Measurability (or operationalizibility)     & Whether measuring the concept being defined is practically possible                                                                                                                                                                                                                                                   & {Yes: \cite{morris_levels_2023} \\No: \cite{bostrom_superintelligence_2014}}                                                                                                                                                                                                                                                                                                                                                           \\
Defined by processes or defined by outcomes & {Defined by processes: definition contains conditions on what the underlying cognitive or neural processes must be like; often these conditions impose human-like properties (e.g. neural structure, consciousness)\\ \\Defined by outcomes: definition only refers to what the system can accomplish, not how it works} & {Defined by processes: \cite{goertzel2012architecture, searle_minds_1980, summerfield_natural_2023, smart_beyond_2015} \\ Defined by outcomes only: \cite{morris_levels_2023}  \\ \\ Additional dimension of difference between outcomes-based views: which outcomes? \\ Maximizing economic value: \cite{openai_2018, suleyman_coming_2023} \\Maximizing a broader set of values: \cite{morris_levels_2023}}                                                                                                                                                      \\
Sociality                                   & Whether measurements of intelligence are carried out in environments where individuals are acting by themselves or where agents interact in a social environment with other agents                                                                                                                                    & {Social definitions: \cite{bostrom_superintelligence_2014, attard-frost_queering_2023} \\Only individualistic measurements: \cite{morris_levels_2023, chollet_measure_2019}} \\

Restricted to instrumental reason           & Whether the definition of intelligence includes the ability to determine what goals the system should pursue, as opposed to only following goals set by some external party                                                                                                                                           & Instrumental only: \cite{legg_collection_2007, bostrom_superintelligence_2014}                                                                                                                                                                                                                                                                                                                                                                                               \\
Which tasks or benchmarks?                     & Which tasks or metrics are included/weighted higher in the definition or measurement of AGI?                                                                                                                                                                                                                          & {Prioritizing language:  \cite{aguera_y_arcas_artificial_2023} . \\Prioritizing logical/mathematical reasoning: \\ \cite{Marcus_2023}}                                                                                                                                                                                                                                                      \\
Generality                                  & How much generality is required to achieve human-level intelligence                                                                                                                                                                                                                                         & {Generality is necessary: \cite{goertzel_artificial_2014, morris_levels_2023, aguera_y_arcas_artificial_2023, summerfield_natural_2023} \\ Generality is not necessary:\\ \cite{bostrom_superintelligence_2014}}                                                                                                                                                                                                                          \\
Use of \emph{g} as an analogy                      & Whether generality is defined by a \emph{g} factor measured analogously to \emph{g} in human intelligence                                                                                                                                                                                                                           & {Uses \emph{g}: \cite{hernandez-orallo_general_2021} \\ Does not use \emph{g}: Most other accounts}                                                                                                                                                                                                                                                                                                                                                        
\end{longtblr}
\end{landscape}

\end{document}